\documentclass[aps,prd,twocolumn,superscriptaddress,amsmath,widetext,amssymbepsfig,showpacs]{revtex4-1}
\usepackage{graphicx}
\usepackage{dcolumn}
\usepackage{bm}
\usepackage{natbib}
\newcommand{\be}{\begin{equation}}
\newcommand{\ee}{\end{equation}}
\newcommand{\bea}{\begin{eqnarray}}
\newcommand{\eea}{\end{eqnarray}}

\newcommand{\Mpc}{{\rm ~Mpc}}

\urldef\mgcamb\url{http://www.sfu.ca/~aha25/MGCAMB.html}

\begin{document}
%opening
\title{Constraints on Modified Gravity from ACT and SPT}

\author {Andrea Marchini}
\affiliation{Physics Department and INFN, Universit\`a di Roma ``La Sapienza'', Ple Aldo Moro 2, 00185, Rome, Italy}

\author {Alessandro Melchiorri}
\affiliation{Physics Department and INFN, Universit\`a di Roma ``La Sapienza'', Ple Aldo Moro 2, 00185, Rome, Italy}

\author {Valentina Salvatelli}
\affiliation{Physics Department and INFN, Universit\`a di Roma ``La Sapienza'', Ple Aldo Moro 2, 00185, Rome, Italy}

\author {Luca Pagano}
\affiliation{Physics Department and INFN, Universit\`a di Roma ``La Sapienza'', Ple Aldo Moro 2, 00185, Rome, Italy}

\begin {abstract}
The Atacama Cosmology Telescope (ACT) and the South Pole Telescope (SPT) have recently provided new and precise measurements of
the Cosmic Microwave Background anisotropy damping tail. This region of the CMB angular spectra, thanks to
the angular distortions produced by gravitational lensing, can probe the growth of matter perturbations and
provide a new test for general relativity.
Here we make use of the ACT and SPT power spectrum measurements (combined with the recent WMAP9 data) to constrain $f(R)$
gravity theories. Adopting a parametrized approach, we obtain an upper limit on the lengthscale of the theory of 
$B_0 < 0.86$ at $95 \%$ c.l. from ACT, while we get a much stronger limit from SPT with 
$B_0 < 0.14$ at $95 \%$ c.l..
\end {abstract}


\maketitle

\section {Introduction} \label {sec:intro}

The major goal of modern cosmology is to understand the source of cosmic acceleration.
One of the possible solutions to this puzzling phenomenon is to modify general relativity on very large scales 
in order to allow an accelerating phase in matter-only universes.
Examples of such "Modified Gravity" (hereafter MG) models are  $f(R)$ theories~\cite{Starobinsky:1980te, Capozziello:2003tk, Carroll:2003wy, Starobinsky:2007hu, Nojiri:2008nk} and in the recent years several authors have searched for modified gravity and departures from general
relativity in cosmological data~\cite{Song:2006ej,Bean:2006up,Pogosian:2007sw,Tsujikawa:2007xu,Zhao:2008bn,Lue:2003ky,Koyama:2005kd, Song:2006jk, Song:2007wd, Cardoso:2007xc, Giannantonio:2008qr,caldwell2007,daniel2008,giannantonio2009,martinelli2010,
daniel2010} 
 
The recent precise measurements of the Cosmic Microwave Background damping tail from the
Atacama Cosmology Telescope (ACT) \cite{act2013} and the South Pole Telescope (SPT) 
\cite{spt2013} are offering a new opportunity to further test MG theories.

The shape of the damping tail of the CMB anisotropies depends strongly from the effect of lensing 
caused by the intervening matter densities along the line of sight of the CMB photons.
CMB lensing therefore probes the growth of perturbations up to redshift $z \sim 6$.
Since the amplitude and the evolution of matter perturbations can be drastically
altered in MG theories a precise detection of the CMB lensing
can place strong constraints on these deviations and possibly identify them
(see e.g. \cite{calmod}).

However the ACT and SPT experiments are reporting quite different constraints
on the amount of CMB lensing(see the discussion in \cite{tail2013}). 
Parametrizing the lensing amplitude by an effective 
amplitude $A_L$, that is $A_L=1$ in case of the standard expected signal
and $A_L=0$ in case of no lensing (see \cite{calamp} for a definition), 
the ACT data provide the constraint $A_L=1.7 \pm 0.38$ at
$68 \%$ c.l. (\cite{act2013}), therefore indicating a larger amplitude, while the SPT is 
more consistent with the standard expectations with $A_L=0.85\pm0.15$, again at $68 \%$ c.l.
(\cite{spt2013}).

The $A_L$ parameter is clearly an effective parameter 
and can be used just to indicate possible deviations from the expectations of the standard scenario.
It is therefore timely, as we do in this paper, 
to analyse the results from ACT and SPT in the context of more physically consistent scenarios,
as MG theories. 

Here we adopt the parametrized modified gravity scenario
presented in \cite{MGCAMB2011}, restricting our analysis to the case of $f(R)$ theories.
In this model, the background expansion is identical to the one produced by
a cosmological constant, while the evolution of perturbation is altered
and depends on a single parameter $B_0$ that represents the 
length-scale of the theory \cite{silvestri}.

Since the ACT and SPT datasets are providing results that are significantly different,
we take a conservative approach to analyse each dataset and discuss the corresponding 
results separately.  The ACT and SPT datasets are combined with the
recent data release from nine of observations from the Wilkinson Microwave
Anisotropy Probe (WMAP9) \cite{wmap9},

The paper is organized as follows. In Section~\ref {sec:theories} we present the modified gravity model considered for our 
analysis, in Sec.~\ref{sec:analysis} we describe the analysis method and in Sec.~\ref {sec:constraints} we present our results. 
We conclude in Section \ref {sec:concl}.

\section {Parametrized $f(R)$ Gravity} \label{sec:theories}

The $f(R)$ theories are currently one of the most popular class of MG models. 
These models generalize the Einstein-Hilbert action replacing the Ricci scalar with a function of $R$ itself.
The generic modified action is 

\begin{equation}
S=\int{d^4x \sqrt{-g} \big[\frac{f(R)}{2k^2}+\mathcal{L}_m]}
\label{action}
\end{equation}

where $k^2=8\pi G$ (c=1) and $\mathcal{L}_m$ is the matter lagrangian density. 

Focusing on this particular MG category is interesting for two main reasons. 
Firstly, their modified Lagrangian is quite simple and generic, since the modified dynamic at every scale is recovered 
using only the first order invariant. Secondly, some models belonging to this class have been shown 
to satisfy both cosmological viability conditions and local tests of gravity, thanks to the chameleon mechanism ~\cite{HS,Starob,AppBatt}.

In order to reproduce the effects of $f(R)$ gravity in the evolution of matter perturbations here we adopt a generic MG parametrized form, 
proposed in \cite{MGCAMB2011}, specializing it to the $f(R)$ case. In this parametrization the background is fixed to that of 
$\Lambda$CDM  and the modifications in the linearized Einstein equation are encoded in two scale- and time-dependent
parametric function $\mu(k,a)$ and $\gamma(k,a)$

\begin{equation}
k^2\Psi = - \mu(k,a) 4 \pi G a^2 \lbrace \rho \Delta + 3(\rho + P) \sigma \rbrace \label{mg-poisson}
\end{equation}
\begin{equation}
k^2[\Phi - \gamma(k,a) \Psi] = \mu(k,a)  12 \pi G a^2   (\rho + P) \sigma \label{mg-anisotropy}
\end{equation}
where $\Psi$ and $\Phi$ are the two scalar metric potentials in the Newtonian gauge, $\sigma$ is the anisotropic stress that vanishes for baryons and CDM but not for relativistic species, $\delta\equiv \delta\rho/\rho$ is the density contrast and $\rho \Delta$ is the comoving density perturbation, defined as

\begin{equation}
\rho \Delta = \rho \delta + 3 \frac{Ha}{k}(\rho +P)v \ 
\end{equation}
where $v$ is the velocity field.

It has been shown in \cite{Bertschinger:2008zb} that we can recover the $f(R)$ theories choosing the following parametric form for $\mu(k,a)$ and $\gamma(k,a)$

%\begin{equation}
%\mu(k,a)=\frac{1+\beta_1\lambda_1^2\,k^2a^s}{1+\lambda_1^2\,k^2a^s} \,, \\
%\label{BZ}
%\gamma(k,a)=\frac{1+\beta_2\lambda_2^2\,k^2a^s}{1+\lambda_2^2\,k^2a^s} \ ,
%\end{equation}

\begin{equation}
\mu(k,a)=\frac{1+\frac{4}{3}\lambda_1^2\,k^2a^s}{1+\lambda_1^2\,k^2a^s} \,, \\
\label{BZ}
\gamma(k,a)=\frac{1+\frac{2}{3}\lambda_1^2\,k^2a^s}{1+\frac{4}{3}\lambda_1^2\,k^2a^s} \ 
\end{equation}

Viable $f(R)$ models must have $s\sim 4$  in order to closely mimic $\Lambda$CDM expansion \cite{silvestri}, that is the case we are interested in.
Indeed the only free parameter we consider in our analysis is the characteristic lengthscale $\lambda_1$. It is usual expressed in literature
in term of the dimensionless parameter $B_0$ as follows:

\begin{equation}\label{B0}
B_0=\frac{2H_0^2\lambda_1^2}{c^2}
\end{equation}

i.e. it gives the lenghtscale in units of the horizon scale.

\section {Data Analysis Method} \label{sec:analysis}

Our theoretical models are computed with the public available code
MGCAMB \cite{MGCAMB2011} v .2 while the analysis is based on a modified version
of CosmoMC \cite{Lewis:2002ah} a Monte Carlo Markov Chain code. 

We consider the following set of recent CMB data (publically available
on the corresponding web pages): WMAP9 \cite{wmap9}, SPT \cite{spt2013}, 
ACT \cite{act2013} including measurements up to a 
maximum multipole number of $l_{\rm max}=3750$.

For the ACT experiment we use the "lite" 
version of the likelihood \cite{dunkleyact} that has been tested to be correct also
in the case of the extension respect to $\Lambda$-CDM models.

We also consider a gaussian prior on the Hubble constant (hereafter HST prior) 
$H_0=73.8\pm2.4 \,\mathrm{km}\,\mathrm{s}^{-1}\,\mathrm{Mpc}^{-1}$, consistently with the measurements of the HST \cite{hst}. 

We include information from 
measurements of baryonic acoustic oscillations (BAO) from galaxy surveys, 
combining four datasets: 6dFGRS from \cite{beutler/etal:2011},
SDSS-DR7 from \cite{padmanabhan/etal:2012}, SDSS-DR9
from \cite{anderson/etal:2012} and WiggleZ from \cite{blake/etal:2012}.
We refer to this dataset as BAO.

We sample a seven-dimensional set of cosmological parameters,
adopting flat priors on them: the $B_0$ modified gravity parameter, the baryon 
and cold dark matter densities $\Omega_{\rm b} h^2$ and $\Omega_{\rm c} h^2$, the ratio of the sound horizon 
to the angular diameter distance at decoupling $\theta$, the optical 
depth to reionization $\tau$, the scalar spectral index $n_s$, the 
overall normalization of the spectrum $A_s$ at $k=0.002\Mpc^{-1}$.

Given the tension between the ACT and SPT experiment in the lensing ampllitude, 
we also consider variations in the lensing amplitude parameter $A_L$
as defined in \cite{calamp}. 
Finally, the amount of helium abundance in the universe $Y_p$ is fixed by assuming
Big Bang Nucleosynthesis in the standard case of three neutrino families.

\section {Results} \label {sec:constraints}

\begin{figure*}[htb!]
\includegraphics[width=6.2cm,angle=-90]{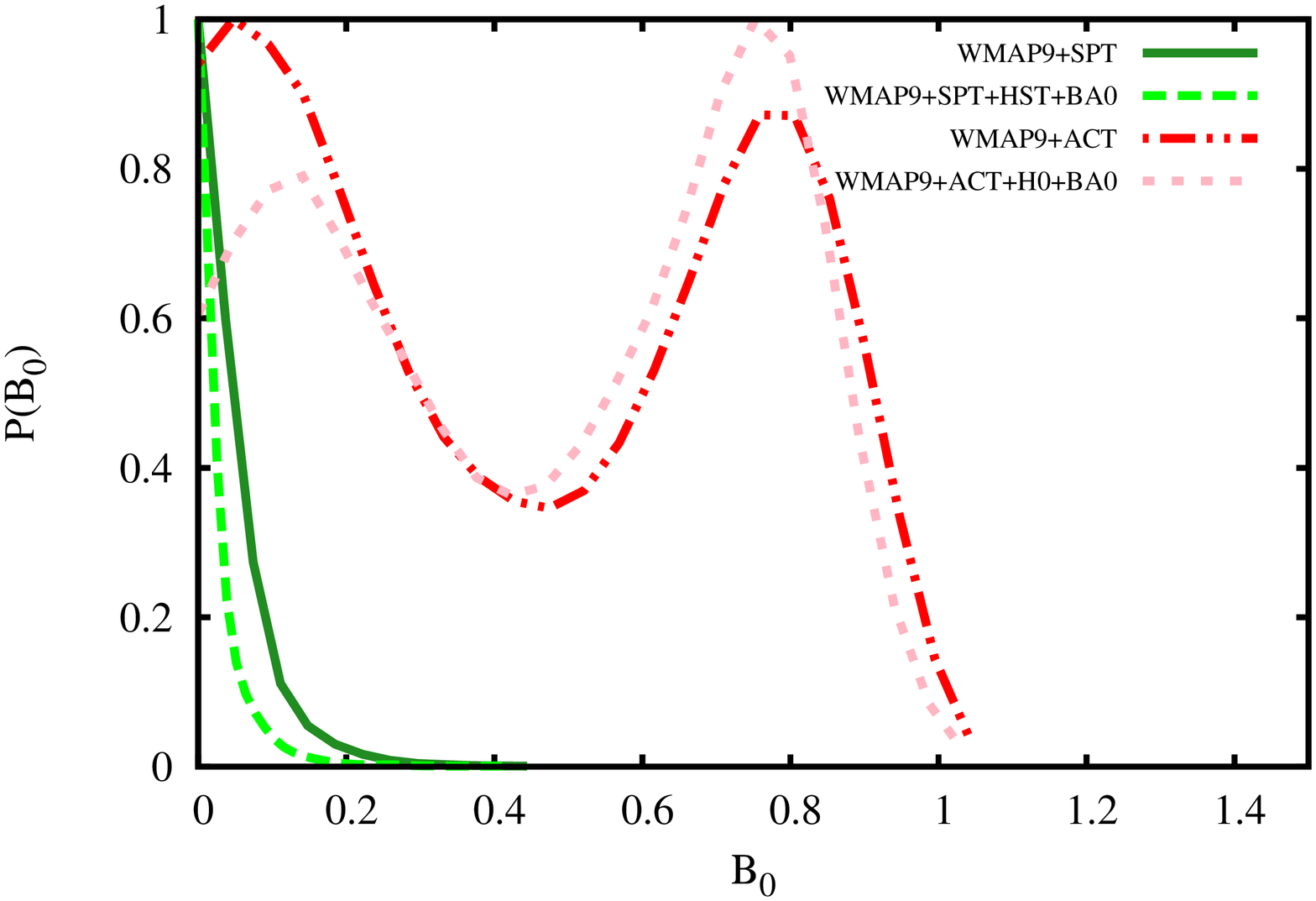}
\includegraphics[width=6.2cm,angle=-90]{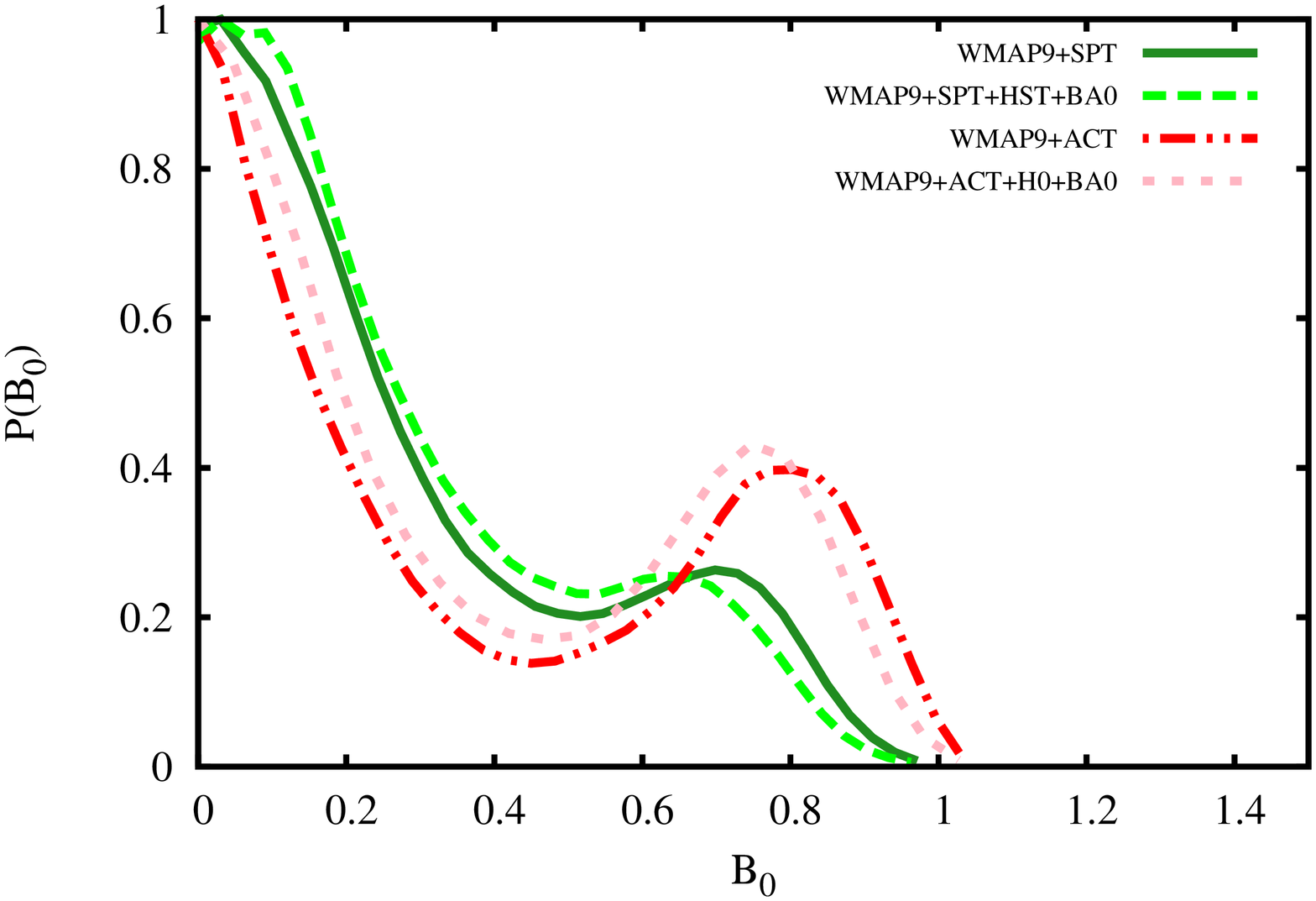}
\caption{Posterior distribution functions for the $B_0$ parameter in the case
of $A_L=1$ (Left panel) and $A_L$ free (Right Panel). The different lensing
amplitude measured by ACT makes MG model more consistent with the data and
a bimodal posterior distribution is present (Left Panel).
When variations in $A_L$ are considered the SPT bound is weaker, while the
ACT dataset is more consistent with GR. The bimodal distribution still present
in the right panel is due to the low WMAP anisotropy at large angular
scales.}
\label{figure}
\end{figure*} 

\begin{figure*}[htb!]
\includegraphics[width=6.2cm,angle=-90]{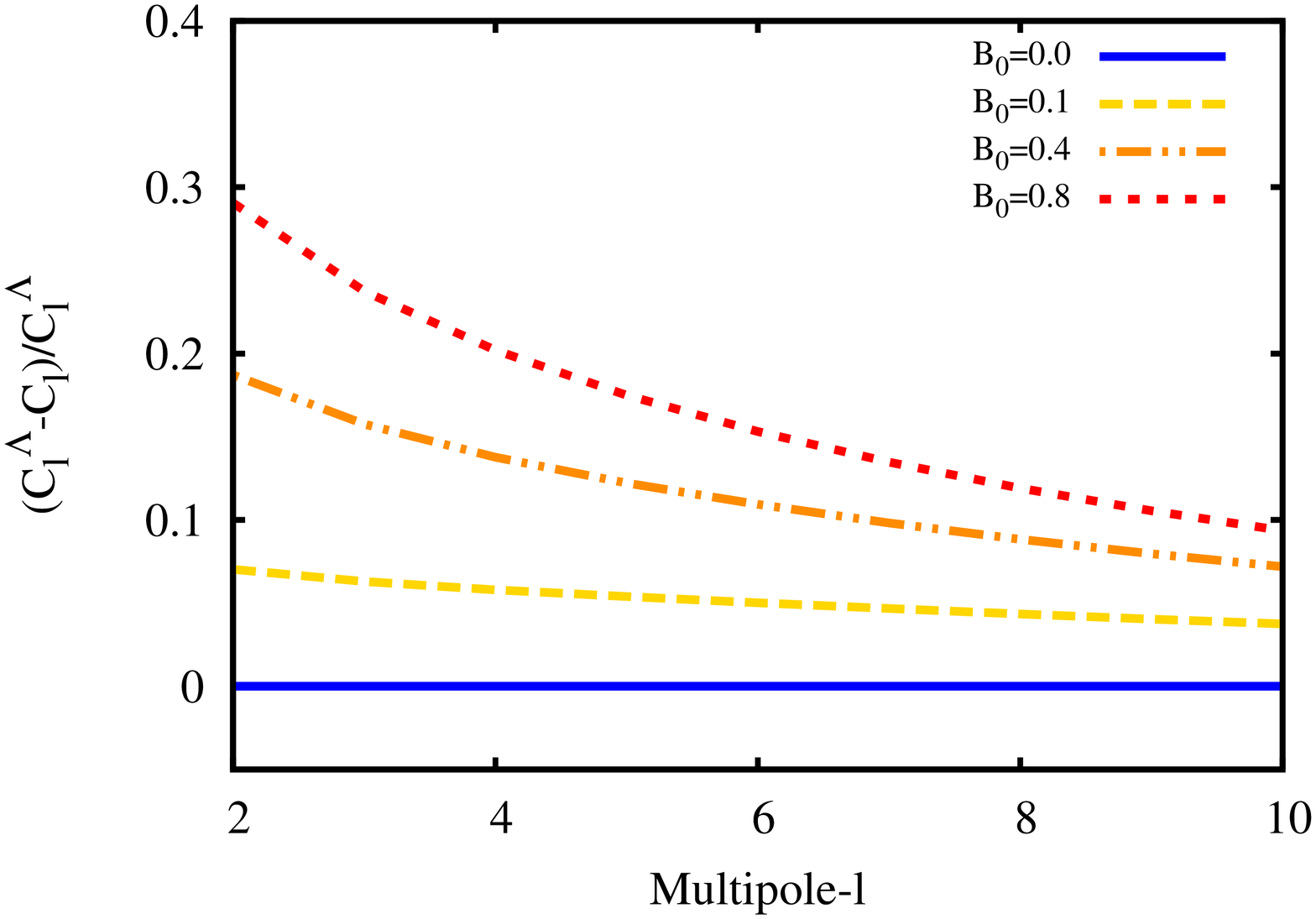}
\includegraphics[width=6.2cm,angle=-90]{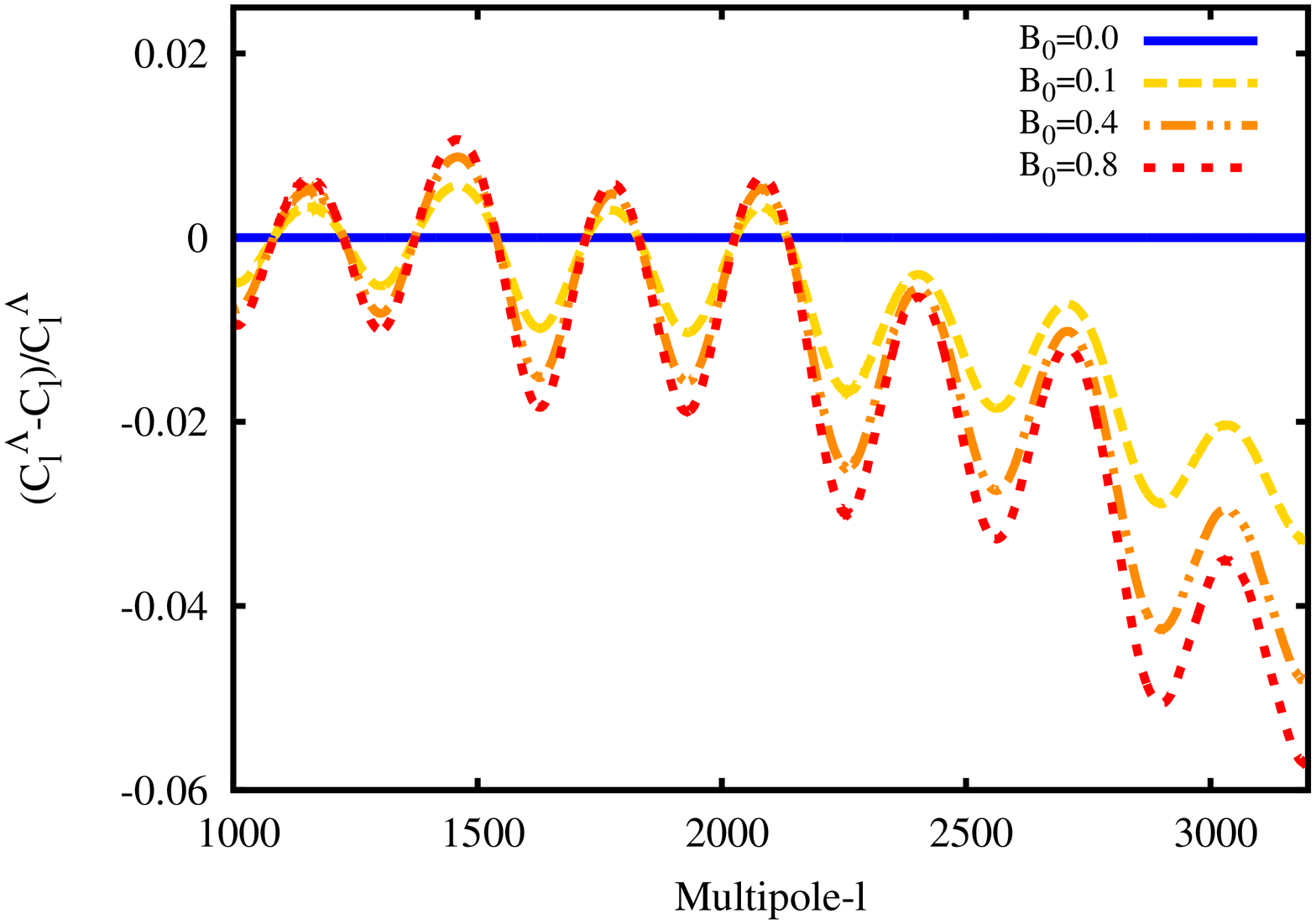}
\caption{Effect of $B_0$ on the CMB angular temperature power spectrum. We plot the differences respect to the
standard $\Lambda$-CDM model. On the left panel we see that the anisotropy at low multipoles decreases
as $B_0$ is increased. A larger $B_0$ is therefore more compatible with the low WMAP quadrupole.
On the right panel we see that the anisotropies on large angular scales are increased respect to
$\Lambda$-CDM. The effect is due to an increase in the CMB lensing amplitude.}
\label{figure2}
\end{figure*}

\begin{table*}[htb!]
\begin{center}
\begin{tabular}{|l||c||c|c||c|}
\hline
\hline
 Parameters & {\bf SPT}+WMAP9 & {\bf ACT}+WMAP9 & {\bf SPT}+WMAP9+HST+BAO & {\bf ACT}+WMAP9+HST+BAO\\
\hline
\hline
$\Omega_b h^2$ & $0.02224\pm0.00034$ & $0.02281\pm0.00044$ & $0.02235\pm0.00033$ & $0.02279\pm0.00040$\\
$\Omega_c h^2$ & $0.1091\pm0.0036$ & $0.1142\pm0.0044$ & $ 0.1119\pm0.0023$ & $0.1149\pm0.0028$\\
$100\theta$ & $1.0428\pm0.0010$ & $1.0402\pm0.0020$ & $1.04257\pm0.00098$ & $1.0403\pm0.0019$\\
$\tau$ & $0.0827\pm0.013$ & $0.091\pm0.014$ & $0.080\pm0.012$ & $0.090\pm0.013$\\
$n_s$ & $0.9676\pm0.0093$ & $0.973\pm0.012$ & $0.9633\pm0.0078$ & $0.9724\pm0.0096$\\
$B_0$ & $<0.14$ (95\% c.l.) & $<0.90$ (95\% c.l.) & $<0.12$ (95\% c.l.) & $<0.86$ (95\% c.l.)\\
$H_0 [\mathrm{km}/\mathrm{s}/\mathrm{Mpc}]$ & $72.2\pm1.7$ & $70.2\pm2.1$ & $70.9\pm1.0$ & $70.0\pm1.3$\\
$\log(10^{10} A_s)$ & $3.060\pm0.027$ & $3.174\pm0.045$ & $3.066\pm0.025$ & $3.185\pm0.035$\\
$\Omega_{\rm \Lambda}$ & $0.747\pm0.018$ & $0.721\pm0.025$ & $0.733\pm 0.012$ & $0.718\pm0.015$\\
$\Omega_{\rm m}$ & $0.253\pm0.018$ & $0.279\pm0.025$ & $0.267\pm0.0012$ & $0.282\pm0.015$\\
Age/Gyr & $13.689\pm0.066$ & $13.71\pm0.10$ & $13.724\pm0.053$ & $13.714\pm0.084$\\
$D_{3000}^{SZ}$ & $4.2\pm2.1$ & ---  & $4.0\pm2.1$ & ---\\
$D_{3000}^{CL}$ & $4.8\pm2.0$ & --- & $4.8\pm2.0$ & ---\\
$D_{3000}^{PS}$ & $20.3\pm2.4$ & --- & $20.5\pm2.3$ & ---\\
$A_{SZ}$ & --- & $0.94\pm0.57$ & --- & $0.91\pm0.56$\\
\hline
$\chi^2_{\rm min}/2$ & $3808.25$ & $3799.09$ & $3811.41$ & $3800.89$ \\
\hline
\hline
\end{tabular}
\caption{Constraints on the MG parameter $B_0$ and the standard cosmological
parameters described in the text from ACT and SPT combined with WMAP9, HST prior
and BAO. We report constraints at $68 \%$ confidence level (bounds on $B_0$ are at $95 \%$ c.l).}
\label{standard}
\end{center}
\end{table*}

Our main results are reported in Table I. Since the ACT and SPT datasets are reporting significantly
different constraints on $B_0$ we consider these two datasets separately. 

As we can see, both ACT and SPT are not providing any evidence for MG.
However, the SPT dataset gives significantly stronger constraints on $B_0$ 
($B_0<0.14$ at $95 \%$ c.l.) respect to those derived by ACT ($B_0<0.90$ at $95 \%$ c.l.).
The difference appears as even more striking in Figure 1 (Left Panel), where we report the two posteriors
on $B_0$ coming from the two experiments: while SPT strongly constrain $B_0$, the posterior
from ACT shows a bimodal distribution, suggesting an higher compatibility with modified gravity models.
The reason of this difference is mainly due to the differenr lensing signal present in the
ACT e SPT TT spectra (see \cite{tail2013}): since $f(R)$ MG models increase the
lensing signal they are more consistent with the larger amplitude of ACT than with
the smaller amplitude of SPT. The best fit value for ACT is indeed $B_0 \sim 0.78$ even if this dataset 
still does not provide any compelling evidence for MG.

The inclusion of the HST prior and of the BAO dataset improves the constraints
($B_0<0.12$ at $95 \%$ c.l. from SPT and $B_0<0.86$ at $95 \%$ c.l. from ACT), however
not in a significant way, clearly showing that most of the constraining power is coming
from the CMB spectrum distortions introduced by gravitational lensing.

It is interesting to consider the impact of MG on the standard cosmological parameters.
As we see from the Table, and as already showed in \cite{MGCAMB2011}, there is little
correlation between $B_0$ and the, standard, six cosmological parameters.
We found that the largest correlations are with scalar spectral index $n_S$ and
amplitude $A_S$. However, these correlations changes in function of $B_0$.
When $B_0<<1$, larger $B_0$ is in more agreement with smaller $n_S$ and larger $A_S$.
When $B_0\sim 1$, larger $B_0$ is in more agreement with larger $n_S$ and smaller $A_S$.

\begin{table*}[htb!]
\begin{center}
\begin{tabular}{|l||c||c|c||c|}
\hline
\hline
 Parameters & {\bf SPT}+WMAP9 & {\bf ACT}+WMAP9 & {\bf SPT}+WMAP9+HST+BAO & {\bf ACT}+WMAP9+HST+BAO\\
\hline
\hline
$\Omega_b h^2$ & $0.02204\pm0.00033$ & $0.02294\pm0.00048$ & $0.02215\pm0.00032$ & $0.02289\pm0.00039$\\
$\Omega_c h^2$ & $0.1154\pm0.0027$ & $0.1132\pm0.0043$ & $ 0.1137\pm0.0025$ & $0.1144\pm0.0027$\\
$\theta$ & $1.04200\pm0.00097$ & $1.0406\pm0.0019$ & $1.04233\pm0.00096$ & $1.0403\pm0.0018$\\
$\tau$ & $0.083\pm0.013$ & $0.090\pm0.014$ & $0.085\pm0.013$ & $0.089\pm0.013$\\
$A_L$ & $0.60\pm0.10$ & $1.25\pm0.29$ & $0.62\pm0.11$ & $1.19\pm0.26$\\
$n_s$ & $0.9561\pm0.0084$ & $0.971\pm0.011$ & $0.9598\pm0.0081$ & $0.9699\pm0.0097$\\
$B_0$ & $<0.73$ (95\% c.l.) & $<0.91$ (95\% c.l.) & $<0.77$ (95\% c.l.) & $<0.85$ (95\% c.l.)\\
$H_0 [\mathrm{km}/\mathrm{s}/\mathrm{Mpc}]$ & $69.1\pm1.2$ & $70.8\pm2.1$ & $70.0\pm1.1$ & $70.2\pm1.2$\\
$\log(10^{10} A_s)$ & $3.083\pm0.026$ & $3.177\pm0.040$ & $3.082\pm0.026$ & $3.183\pm0.034$\\
$\Omega_{\rm \Lambda}$ & $0.712\pm0.015$ & $0.727\pm0.024$ & $0.722\pm 0.013$ & $0.721\pm0.014$\\
$\Omega_{\rm m}$ & $0.288\pm0.015$ & $0.273\pm0.024$ & $0.278\pm0.0013$ & $0.279\pm0.014$\\
Age/Gyr & $13.790\pm0.055$ & $13.68\pm0.10$ & $13.760\pm0.053$ & $13.702\pm0.079$\\
$D_{3000}^{SZ}$ & $5.1\pm2.3$ & ---  & $5.0\pm2.3$ & ---\\
$D_{3000}^{CL}$ & $5.2\pm2.1$ & --- & $5.2\pm2.1$ & ---\\
$D_{3000}^{PS}$ & $20.0\pm2.4$ & --- & $20.1\pm2.4$ & ---\\
$A_{SZ}$ & --- & $1.9\pm1.3$ & --- & $1.7\pm1.2$\\
\hline
$\chi^2_{\rm min}/2$ & $3807.34$ & $3798.95$ & $3808.98$ & $3800.64$ \\
\hline
\hline
\end{tabular}
\caption{Constraints on the $f(R)$ parameter $B_0$, the lensing amplitude $A_L$  and the standard cosmological
parameters described in the text from ACT and SPT combined with WMAP9, HST prior
and BAO. We report constraints at $68 \%$ confidence level (bounds on $B_0$ are at $95 \%$ c.l).}
\label{alvariable}
\end{center}
\end{table*}

In order to further test the importance of the lensing signal in constraining modified gravity models
we have performed an analysis by letting variations in the lensing amplitude $A_L$.
The results are reported in Table 2. As we can see, the effect of marginalizing over the lensing amplitude
is clearly to make weaker the SPT constraint and to leave as unaffected the ACT constraint.
However the bimodal distribution present in the ACT case is now suppressed as we can see
from Figure 1 (Right Panel) where we plot the posterior distribution functions for $B_0$. Moreover,
as we can see from the results in Table 2, the ACT lensing signal is consistent with $A_L=1$ in the case of MG
models. We can therefore conclude that while the ACT data does
not show any evidence for MG, the lensing signal is in better agreement with the
$A_L=1$ case in the framework of MG.

It is interesting to note that, while now suppressed from the previous case, the
bimodal distribution is still present when $A_L$ varies and it is also now evident in the SPT dataset.
The reason is that MG $f(R)$ gravity models produce also a lower quadrupole and lower
$\ell$ temperature anisotropy in agreement with the WMAP data
(see Figure 2, left panel).

\section {Conclusions} \label {sec:concl}

In this brief paper we have presented new constraints on $f(R)$ MG models from 
the new recent measurements of the CMB damping tail provided by the ACT and
SPT experiments.
We have found that both experiments show no evidence for deviations from
GR. However, while the SPT data significantly improves the previous constraints
obtained from similar analysis, the ACT data gives much weaker constraints and
shows a bimodal posterior distribution for $B_0$.
We have attributed this different behaviour to the different amplitude of the
lensing signal detected by those experiments and showed that when the lensing
amplitude $A_L$ is let to vary both datasets provide similar constraints.
When $A_L$ is varied, we have found that the ACT data does not show any indication
for $A_L>1$ in the framework of MG models.
Moreover, a bimodal distribution for $B_0$ is present in both ACT and SPT datasets
when we marginalize over $A_L$. This is due to the large angular scale
regime of the measured CMB spectrum, that prefers a low quadrupole and a bluer
spectral index (see e.g. \cite{pandolfi}).

Presenting combined results from ACT and SPT spectra as in \cite{calabrese013} needs
to be done with great care: while compatible in between two standard deviations 
in the case of a standard six parameter
analysis, the two experiments could show very different constraints in extended theoretical frameworks,
especially when the lensing signal plays a significant constraining role.

It is useful to compare the SPT results with previous limits on $B_0$ present in the literature. 
The constrain we obtain from WMAP9+SPT+H0+BAO, in the case $A_L=1$, is much tighter than 
the constraint of $B_0<0.42$ (95\% C.L.) obtained from a combined analysis of cosmological data and
Integrated Sachs Wolfe data \cite{Giannantonio:2009} and from the similar constraints 
$B_0<0.4$ (95\% C.L.) obtained in \cite{MGCAMB2011}  and $B_0<0.42$ (95\% C.L.) from \cite{Lombriser:2012}, where the datasets considered are
slightly different between papers. In \cite{Lombriser:2012} they also found a very tight constrain combining with cluster abundance data ($B_0<0.001$ 95\% C.L.), however this constraint is obtained in the non linear perturbation regime where the simple
treatment of $f(R)$ models we adopt here may not be sufficient.

While the SPT provides a much better constraint, one should however also consider it with great caution, 
given the tension on the lensing amplitude with the ACT dataset.

Finally, the ACT collaboration has provided a determination of the lensing amplitude also
from the four points CMB correlation function (see \cite{act2013}). This amplitude is perfectly consistent with
the standard case, however we prefer here to do not include this dataset for the following 
conservative reasons: a) we prefer to compare the ACT and SPT datasets at the same power spectrum level;
b) the ACT constraint from higher correlations comes from about $50 \%$ of the 
data used in the estimation for the power spectrum (the ACT-E dataset).

The current measurements of CMB lensing will be dramatically improved by the Planck satellite mission,
expected to release new data by end on March 2013.

\subsection*{Acknowledgements}

It is a pleasure to thank Francesco De Bernardis, Eleonora Di Valentino, Massimiliano Lattanzi and Najla Said for help.

\label{lastpage}

\begin{thebibliography}{60}
\expandafter\ifx\csname natexlab\endcsname\relax\def\natexlab#1{#1}\fi
\expandafter\ifx\csname bibnamefont\endcsname\relax
  \def\bibnamefont#1{#1}\fi
\expandafter\ifx\csname bibfnamefont\endcsname\relax
  \def\bibfnamefont#1{#1}\fi
\expandafter\ifx\csname citenamefont\endcsname\relax
  \def\citenamefont#1{#1}\fi
\expandafter\ifx\csname url\endcsname\relax
  \def\url#1{\texttt{#1}}\fi
\expandafter\ifx\csname urlprefix\endcsname\relax\def\urlprefix{URL }\fi
\providecommand{\bibinfo}[2]{#2}
\providecommand{\eprint}[2][]{\url{#2}}

\bibitem[{\citenamefont{Starobinsky}(1980)}]{Starobinsky:1980te}
\bibinfo{author}{\bibfnamefont{A.~A.} \bibnamefont{Starobinsky}},
  \bibinfo{journal}{Phys. Lett.} \textbf{\bibinfo{volume}{B91}},
  \bibinfo{pages}{99} (\bibinfo{year}{1980}).

\bibitem[{\citenamefont{Capozziello et~al.}(2003)\citenamefont{Capozziello,
  Carloni, and Troisi}}]{Capozziello:2003tk}
\bibinfo{author}{\bibfnamefont{S.}~\bibnamefont{Capozziello}},
  \bibinfo{author}{\bibfnamefont{S.}~\bibnamefont{Carloni}}, \bibnamefont{and}
  \bibinfo{author}{\bibfnamefont{A.}~\bibnamefont{Troisi}},
  \bibinfo{journal}{Recent Res. Dev. Astron. Astrophys.}
  \textbf{\bibinfo{volume}{1}}, \bibinfo{pages}{625} (\bibinfo{year}{2003}),
  \eprint{astro-ph/0303041}.

\bibitem[{\citenamefont{Carroll et~al.}(2004)\citenamefont{Carroll, Duvvuri,
  Trodden, and Turner}}]{Carroll:2003wy}
\bibinfo{author}{\bibfnamefont{S.~M.} \bibnamefont{Carroll}},
  \bibinfo{author}{\bibfnamefont{V.}~\bibnamefont{Duvvuri}},
  \bibinfo{author}{\bibfnamefont{M.}~\bibnamefont{Trodden}}, \bibnamefont{and}
  \bibinfo{author}{\bibfnamefont{M.~S.} \bibnamefont{Turner}},
  \bibinfo{journal}{Phys. Rev.} \textbf{\bibinfo{volume}{D70}},
  \bibinfo{pages}{043528} (\bibinfo{year}{2004}), \eprint{astro-ph/0306438}.

\bibitem[{\citenamefont{Starobinsky}(2007)}]{Starobinsky:2007hu}
\bibinfo{author}{\bibfnamefont{A.~A.} \bibnamefont{Starobinsky}},
  \bibinfo{journal}{JETP Lett.} \textbf{\bibinfo{volume}{86}},
  \bibinfo{pages}{157} (\bibinfo{year}{2007}), \eprint{0706.2041}.

\bibitem[{\citenamefont{Nojiri and Odintsov}(2008)}]{Nojiri:2008nk}
\bibinfo{author}{\bibfnamefont{S.}~\bibnamefont{Nojiri}} \bibnamefont{and}
  \bibinfo{author}{\bibfnamefont{S.~D.} \bibnamefont{Odintsov}}
  (\bibinfo{year}{2008}), \eprint{0801.4843}.


\bibitem[{\citenamefont{Song et~al.}(2007{\natexlab{a}})\citenamefont{Song, Hu,
  and Sawicki}}]{Song:2006ej}
\bibinfo{author}{\bibfnamefont{Y.-S.} \bibnamefont{Song}},
  \bibinfo{author}{\bibfnamefont{W.}~\bibnamefont{Hu}}, \bibnamefont{and}
  \bibinfo{author}{\bibfnamefont{I.}~\bibnamefont{Sawicki}},
  \bibinfo{journal}{Phys. Rev.} \textbf{\bibinfo{volume}{D75}},
  \bibinfo{pages}{044004} (\bibinfo{year}{2007}{\natexlab{a}}),
  \eprint{astro-ph/0610532}.

\bibitem[{\citenamefont{Bean et~al.}(2007)\citenamefont{Bean, Bernat, Pogosian,
  Silvestri, and Trodden}}]{Bean:2006up}
\bibinfo{author}{\bibfnamefont{R.}~\bibnamefont{Bean}},
  \bibinfo{author}{\bibfnamefont{D.}~\bibnamefont{Bernat}},
  \bibinfo{author}{\bibfnamefont{L.}~\bibnamefont{Pogosian}},
  \bibinfo{author}{\bibfnamefont{A.}~\bibnamefont{Silvestri}},
  \bibnamefont{and} \bibinfo{author}{\bibfnamefont{M.}~\bibnamefont{Trodden}},
  \bibinfo{journal}{Phys. Rev.} \textbf{\bibinfo{volume}{D75}},
  \bibinfo{pages}{064020} (\bibinfo{year}{2007}), \eprint{astro-ph/0611321}.

\bibitem[{\citenamefont{Pogosian and Silvestri}(2008)}]{Pogosian:2007sw}
\bibinfo{author}{\bibfnamefont{L.}~\bibnamefont{Pogosian}} \bibnamefont{and}
  \bibinfo{author}{\bibfnamefont{A.}~\bibnamefont{Silvestri}},
  \bibinfo{journal}{Phys. Rev.} \textbf{\bibinfo{volume}{D77}},
  \bibinfo{pages}{023503} (\bibinfo{year}{2008}), \eprint{0709.0296}.

\bibitem[{\citenamefont{Tsujikawa}(2008)}]{Tsujikawa:2007xu}
\bibinfo{author}{\bibfnamefont{S.}~\bibnamefont{Tsujikawa}},
  \bibinfo{journal}{Phys. Rev.} \textbf{\bibinfo{volume}{D77}},
  \bibinfo{pages}{023507} (\bibinfo{year}{2008}), \eprint{0709.1391}.

\bibitem[{\citenamefont{Zhao et~al.}(2009{\natexlab{a}})\citenamefont{Zhao,
  Pogosian, Silvestri, and Zylberberg}}]{Zhao:2008bn}
\bibinfo{author}{\bibfnamefont{G.-B.} \bibnamefont{Zhao}},
  \bibinfo{author}{\bibfnamefont{L.}~\bibnamefont{Pogosian}},
  \bibinfo{author}{\bibfnamefont{A.}~\bibnamefont{Silvestri}},
  \bibnamefont{and}
  \bibinfo{author}{\bibfnamefont{J.}~\bibnamefont{Zylberberg}},
  \bibinfo{journal}{Phys. Rev.} \textbf{\bibinfo{volume}{D79}},
  \bibinfo{pages}{083513} (\bibinfo{year}{2009}{\natexlab{a}}),
  \eprint{0809.3791}.

\bibitem[{\citenamefont{Lue et~al.}(2004)\citenamefont{Lue, Scoccimarro, and
  Starkman}}]{Lue:2003ky}
\bibinfo{author}{\bibfnamefont{A.}~\bibnamefont{Lue}},
  \bibinfo{author}{\bibfnamefont{R.}~\bibnamefont{Scoccimarro}},
  \bibnamefont{and} \bibinfo{author}{\bibfnamefont{G.}~\bibnamefont{Starkman}},
  \bibinfo{journal}{Phys. Rev.} \textbf{\bibinfo{volume}{D69}},
  \bibinfo{pages}{044005} (\bibinfo{year}{2004}), \eprint{astro-ph/0307034}.

\bibitem[{\citenamefont{Koyama and Maartens}(2006)}]{Koyama:2005kd}
\bibinfo{author}{\bibfnamefont{K.}~\bibnamefont{Koyama}} \bibnamefont{and}
  \bibinfo{author}{\bibfnamefont{R.}~\bibnamefont{Maartens}},
  \bibinfo{journal}{JCAP} \textbf{\bibinfo{volume}{0601}}, \bibinfo{pages}{016}
  (\bibinfo{year}{2006}), \eprint{astro-ph/0511634}.

\bibitem[{\citenamefont{Song et~al.}(2007{\natexlab{b}})\citenamefont{Song,
  Sawicki, and Hu}}]{Song:2006jk}
\bibinfo{author}{\bibfnamefont{Y.-S.} \bibnamefont{Song}},
  \bibinfo{author}{\bibfnamefont{I.}~\bibnamefont{Sawicki}}, \bibnamefont{and}
  \bibinfo{author}{\bibfnamefont{W.}~\bibnamefont{Hu}}, \bibinfo{journal}{Phys.
  Rev.} \textbf{\bibinfo{volume}{D75}}, \bibinfo{pages}{064003}
  (\bibinfo{year}{2007}{\natexlab{b}}), \eprint{astro-ph/0606286}.

\bibitem[{\citenamefont{Song}(2008)}]{Song:2007wd}
\bibinfo{author}{\bibfnamefont{Y.-S.} \bibnamefont{Song}},
  \bibinfo{journal}{Phys. Rev.} \textbf{\bibinfo{volume}{D77}},
  \bibinfo{pages}{124031} (\bibinfo{year}{2008}), \eprint{0711.2513}.

\bibitem[{\citenamefont{Cardoso et~al.}(2008)\citenamefont{Cardoso, Koyama,
  Seahra, and Silva}}]{Cardoso:2007xc}
\bibinfo{author}{\bibfnamefont{A.}~\bibnamefont{Cardoso}},
  \bibinfo{author}{\bibfnamefont{K.}~\bibnamefont{Koyama}},
  \bibinfo{author}{\bibfnamefont{S.~S.} \bibnamefont{Seahra}},
  \bibnamefont{and} \bibinfo{author}{\bibfnamefont{F.~P.} \bibnamefont{Silva}},
  \bibinfo{journal}{Phys. Rev.} \textbf{\bibinfo{volume}{D77}},
  \bibinfo{pages}{083512} (\bibinfo{year}{2008}), \eprint{0711.2563}.

\bibitem[{\citenamefont{Giannantonio
  et~al.}(2008{\natexlab{a}})\citenamefont{Giannantonio, Song, and
  Koyama}}]{Giannantonio:2008qr}
\bibinfo{author}{\bibfnamefont{T.}~\bibnamefont{Giannantonio}},
  \bibinfo{author}{\bibfnamefont{Y.-S.} \bibnamefont{Song}}, \bibnamefont{and}
  \bibinfo{author}{\bibfnamefont{K.}~\bibnamefont{Koyama}},
  \bibinfo{journal}{Phys. Rev.} \textbf{\bibinfo{volume}{D78}},
  \bibinfo{pages}{044017} (\bibinfo{year}{2008}{\natexlab{a}}),
  \eprint{0803.2238}.

\bibitem{caldwell2007}
  R.~Caldwell, A.~Cooray and A.~Melchiorri,
  %``Constraints on a New Post-General Relativity Cosmological Parameter,''
  Phys.\ Rev.\ D {\bf 76} (2007) 023507
  [astro-ph/0703375 [ASTRO-PH]].
  %%CITATION = ASTRO-PH/0703375;%%

\bibitem{daniel2008}
  S.~F.~Daniel, R.~R.~Caldwell, A.~Cooray and A.~Melchiorri,
  %``Large Scale Structure as a Probe of Gravitational Slip,''
  Phys.\ Rev.\ D {\bf 77} (2008) 103513
  [arXiv:0802.1068 [astro-ph]].
  %%CITATION = ARXIV:0802.1068;%%

\bibitem{giannantonio2009}
  T.~Giannantonio, M.~Martinelli, A.~Silvestri and A.~Melchiorri,
  %``New constraints on parametrised modified gravity from correlations of the CMB with large scale structure,''
  JCAP {\bf 1004} (2010) 030
  [arXiv:0909.2045 [astro-ph.CO]].
  %%CITATION = ARXIV:0909.2045;%%

\bibitem{martinelli2010}
  M.~Martinelli, E.~Calabrese, F.~De Bernardis, A.~Melchiorri, L.~Pagano and R.~Scaramella,
  %``Constraining Modified Gravity with Euclid,''
  Phys.\ Rev.\ D {\bf 83} (2011) 023012
  [arXiv:1010.5755 [astro-ph.CO]].
  %%CITATION = ARXIV:1010.5755;%%

\bibitem{daniel2010}
  S.~F.~Daniel, E.~V.~Linder, T.~L.~Smith, R.~R.~Caldwell, A.~Cooray, A.~Leauthaud and L.~Lombriser,
  %``Testing General Relativity with Current Cosmological Data,''
  Phys.\ Rev.\ D {\bf 81} (2010) 123508
  [arXiv:1002.1962 [astro-ph.CO]].
  %%CITATION = ARXIV:1002.1962;%%

\bibitem{act2013} 
  J.~L.~Sievers, R.~A.~Hlozek, M.~R.~Nolta, V.~Acquaviva, G.~E.~Addison, P.~A.~R.~Ade, P.~Aguirre and M.~Amiri {\it et al.},
  %``The Atacama Cosmology Telescope: Cosmological parameters from three seasons of data,''
  arXiv:1301.0824 [astro-ph.CO].
  %%CITATION = ARXIV:1301.0824;%%

\bibitem{spt2013} 
 Z.~Hou, C.~L.~Reichardt, K.~T.~Story, B.~Follin, R.~Keisler, K.~A.~Aird, B.~A.~Benson and L.~E.~Bleem {\it et al.},
  %``Constraints on Cosmology from the Cosmic Microwave Background Power Spectrum of the 2500-square degree SPT-SZ Survey,''
  arXiv:1212.6267 [astro-ph.CO].
  %%CITATION = ARXIV:1212.6267;%%

\bibitem{tail2013}
  E.~Di Valentino, S.~Galli, M.~Lattanzi, A.~Melchiorri, P.~Natoli, L.~Pagano and N.~Said,
  %``Tickling the CMB damping tail: scrutinizing the tension between the ACT and SPT experiments,''
  arXiv:1301.7343 [astro-ph.CO].
  %%CITATION = ARXIV:1301.7343;%%

\bibitem{calmod}
  E.~Calabrese, A.~Cooray, M.~Martinelli, A.~Melchiorri, L.~Pagano, A.~Slosar and G.~F.~Smoot,
  %``CMB Lensing Constraints on Dark Energy and Modified Gravity Scenarios,''
  Phys.\ Rev.\ D {\bf 80} (2009) 103516
  [arXiv:0908.1585 [astro-ph.CO]].
  %%CITATION = ARXIV:0908.1585;%%

\bibitem{calamp}
  E.~Calabrese, A.~Slosar, A.~Melchiorri, G.~F.~Smoot and O.~Zahn,
  %``Cosmic Microwave Weak lensing data as a test for the dark universe,''
  Phys.\ Rev.\ D {\bf 77} (2008) 123531
  [arXiv:0803.2309 [astro-ph]].
  %%CITATION = ARXIV:0803.2309;%%

\bibitem{silvestri}
G.-B. Zhao, L. Pogosian, A. Silvestri, and J. Zylberberg,
%"Searching for modified growth patterns with tomographic surveys"
Phys. Rev. D79, 083513 (2009), 0809.3791

\bibitem{wmap9}
  G.~Hinshaw, D.~Larson, E.~Komatsu, D.~N.~Spergel, C.~L.~Bennett, J.~Dunkley, M.~R.~Nolta and M.~Halpern {\it et al.},
  %``Nine-Year Wilkinson Microwave Anisotropy Probe (WMAP) Observations: Cosmological Parameter Results,''
  arXiv:1212.5226 [astro-ph.CO].
  %%CITATION = ARXIV:1212.5226;%%


\bibitem{Lewis:2002ah}
  A.~Lewis and S.~Bridle,
  Phys.\ Rev.\ D {\bf 66}, 103511 (2002)
  [arXiv:astro-ph/0205436].


\bibitem{beutler/etal:2011}
{Beutler}, F., et~al., Montly Notices of the Royal Astronomical Society, 416, 3017, 2011

\bibitem{padmanabhan/etal:2012}
{Padmanabhan}, N., {Xu}, X., {Eisenstein}, D.~J., {Scalzo}, R., {Cuesta},
  A.~J., {Mehta}, K.~T., \& {Kazin}, E. 2012, ArXiv e-prints, arXiv:1202.0090

\bibitem{pandolfi}
 S.~Pandolfi, E.~Giusarma, M.~Lattanzi and A.~Melchiorri,
  %``Inflation with primordial broken power law spectrum as an alternative to the concordance cosmological model,''
  Phys.\ Rev.\ D {\bf 81} (2010) 103007.
  %%CITATION = PHRVA,D81,103007;%%

\bibitem{hst}
A.~G.~Riess, L.~Macri, S.~Casertano, H.~Lampeitl, H.~C.~Ferguson, A.~V.~Filippenko, S.~W.~Jha and W.~Li {\it et al.},
  %``A 3% Solution: Determination of the Hubble Constant with the Hubble Space Telescope and Wide Field Camera 3,''
  Astrophys.\ J.\  {\bf 730}, 119 (2011)
  [Erratum-ibid.\  {\bf 732}, 129 (2011)]
  [arXiv:1103.2976 [astro-ph.CO]].
  %%CITATION = ARXIV:1103.2976;%%


\bibitem{anderson/etal:2012}
L. Anderson, et~al. 2012, ArXiv e-prints, arXiv:1203.6594

\bibitem{blake/etal:2012}
C. Blake et al., Montly Notices of the Royal Astronomical Society, 425, 405, 2012 

\bibitem{dunkleyact} 
  J.~Dunkley, E.~Calabrese, J.~Sievers, G.~E.~Addison, N.~Battaglia, E.~S.~Battistelli, J.~R.~Bond and S.~Das {\it et al.},
  %``The Atacama Cosmology Telescope: likelihood for small-scale CMB data,''
  arXiv:1301.0776 [astro-ph.CO].
  %%CITATION = ARXIV:1301.0776;%%
\bibitem{HS}
W.~Hu, I.~Sawicki,
%'Modelof f(R) cosmic acceleration that evade solar system tests'
Phys.\ Rev.\ D {\bf 76}, 064004 (2007)

\bibitem{Starob}
A.A.~Starobinsky,
%'Diseappearing cosmological constant in f(R) gravity'
JETP \ Lett {\bf 86}, 157-163 (2007)

\bibitem{AppBatt}
S.~Appleby, R.A.~Battye,
%'Do consistent f(R) models mimic general relativity plus LAMBDA?'
Phys.\ Lett.\ B {\bf 654},101016 (2007)

\bibitem{MGCAMB2011}
A.~Hojjati, L.~Pogosian, G.~Zhao,
%'Testing gravity with CAMB and CosmoMC',
JCAP {\bf 005}, 1108 (2011)

\bibitem{Bertschinger:2008zb}
  E.~Bertschinger and P.~Zukin,
  %``Distinguishing Modified Gravity from Dark Energy,''
  Phys.\ Rev.\  D {\bf 78}, 024015 (2008)
  %[arXiv:0801.2431 [astro-ph]].
  
\bibitem{Giannantonio:2009}
  T.~Giannantonio, M.~Martinelli, A.~Silvestri and A.~Melchiorri,
  %``New constraints on parametrized modified gravity from correlations of the
  %CMB with large scale structure,''
  JCAP {\bf 1004}, 030 (2010)
  %[arXiv:0909.2045 [astro-ph.CO]].
  %%CITATION = JCAPA,1004,030;%%   
  
\bibitem{Lombriser:2012}
  L.~Lombriser, A.~Slosar, U.~Seljak, W.~Hu,
  %`Constraints on f(R) gravity from probing the large-scale structure'
  Phys.\ Rev.\  D {\bf 85}, 124038 (2012)

\bibitem{calabrese013}
E.~Calabrese, R.~ée A.~Hlozek, N.~Battaglia, E.~S.~Battistelli, J.~R.~Bond, J.~Chluba, D.~Crichton and S.~Das {\it et al.},
  %``Cosmological Parameters from Pre-Planck CMB Measurements,''
  arXiv:1302.1841 [astro-ph.CO].
  %%CITATION = ARXIV:1302.1841;%%
\end{thebibliography}
\end{document}